\documentclass{paper}
\usepackage{epsfig, graphicx}
\usepackage[english]{babel}

\begin{document}

\title{The Critical Crossover at the n-Hexane –- Water Interface}
\author{Aleksey M. Tikhonov\/\thanks{tikhonov@kapitza.ras.ru}}
\maketitle

\leftline{\it Kapitza Institute for Physical Problems, Russian Academy of Sciences,}
\leftline{\it ul. Kosygina 2, Moscow, 119334, Russia}

\rightline{\today}

\abstract{According to estimates of the parameters of the critical crossover in monolayers of long-chain alcohol molecules adsorbed at the n-hexane –- water interface, all systems in which this phenomenon is
observed are characterized by the same value of the critical exponent $\nu\approx 1.8$.}
\vspace{0.25in}

\large
Atoms or molecules adsorbed on the surface of a liquid or crystal frequently form a spatially inhomogeneous structure in which domains of two homogeneous phases coexist [1–-4].
Both of these phases tend to intermixing, since the formation of one-dimensional
interphase boundaries leads to a significant decrease in the system energy [5].
An evident consequence of this is the impossibility of a two-dimensional (2D) first-order phase transition in this system; instead, an infinite sequence of phase transitions
(critical crossover) must take place [6].

This article presents the results of an analysis of
experimental data obtained earlier [7, 8], which
allowed a critical parameter of the crossover at the n-hexane–water interface to be established.

A macroscopically flat interphase boundary (interface) between n-hexane (a nonpolar organic solvent)
and water (see Fig. 1) offers an example of the system,
featuring the phenomenon of critical crossover. Under
normal conditions, n-hexane (saturated hydrocarbon
with the formula C$_6$H$_{14}$, a density of $\sim$ 0.65 g/cm$^3$ at
T = 298 K, and a boiling temperature of about 342 K)
and water exhibit virtually no mutual solubility.

It was reported earlier [7–-9] that a surface electrical double layer can form at the n-hexane -– water interface owing to adsorption (from hexane solutions) of long-chain molecules of fluorinated alcohols, such as 1,1,2,2-tetrahydroheptadecafluorodecanol (FC$_{10}$OH) and 1,1,2,2-tetrahydrohenicosafluorododecanol (FC$_{12}$OH),
or normal alkanols, such as n-tetracosanol (C$_{24}$OH) and n-triacontanol (C$_{30}$OH).
The fluorocarbon chain of FC$_{12}$OH is longer by two –-CF$_2$-– units than that of FC$_{10}$OH, and the hydrocarbon chain of C$_{30}$OH is longer by six –-CH$_2$–- units than that of C$_{24}$OH. The main difference between the properties of fluorocarbon and hydrocarbon chain molecules is their flexibility. Indeed, the former molecules at room temperature can be considered absolutely rigid rods, whereas the latter are susceptible to conformation isomerization.

Synchrotron radiation reflectometry data show
that molecules of the aforementioned substances at
sufficiently low temperatures adsorb from a solution in
liquid hydrocarbon at the n-hexane–-water interface in
the form of a monolayer (Gibbs monolayer) with a
certain set of thermodynamc parameters $(p, T, c)$. It
was found that the long-chain molecules of various
alcohols are ordered differently on the water surface.
The density of molecules in the condensed low-temperature phase of monolayers of fluorinated alcohols is close to the density of the corresponding volume crystals, whereas the density of the condensed phase of saturated monatomic alcohols is close to that of the
high-molecular-mass hydrocarbon liquid.
These alcohols are nearly insoluble in water at room temperature.

\begin{figure}
\hspace{1.0in}
\epsfig{file=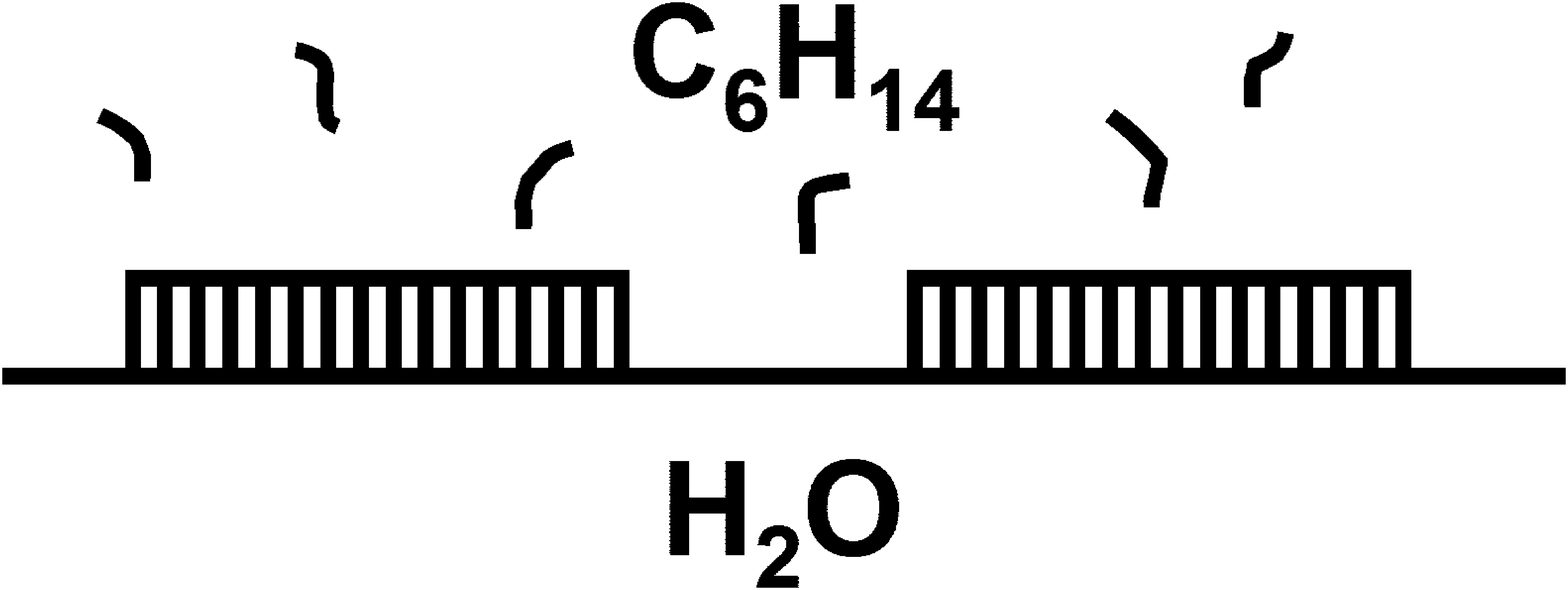, width=0.5\textwidth}

Figure 1. Schematic diagram of a macroscopically flat interface between n-hexane
(a light nonpolar organic solvent) and water, which is oriented by the gravitational field.
Under normal conditions, n-hexane and water are immiscible liquids. In a broad temperature range, domains of the condensed low-temperature phase in the monolayer of adsorbed alcohol
(vertical rectangles) coexist with domains of the gaseous high-temperature phase and both
occur in equilibrium with the volume phase of the organic solvent, which serves as a reservoir for
surfactant molecules.
\end{figure}

As the temperature $T$ is increased at normal pressure ($p = 1$ bar), the monolayer exhibits a transition from the condensed to the gaseous phase. The transition temperature $T_c$ is determined by the alcohol concentration $c$ in the volume of the organic solvent
(a reservoir for surfactant molecules). By analogy with
3D systems, we can say that the monolayer of a fluorinated alcohol exhibits a solid–-gas transition, while the monolayer of a saturated monatomic alcohol exhibits
a transition from liquid (amorphous monolayer) to
gas.

However, there are several experimental facts evidence the existence of a spatially inhomogeneous
equilibrium structure on the surface in the vicinity of
$T_c$ [2, 7]. At fixed p and c, the surface exhibits coexistence of the domains of condensed and gaseous phases in a rather wide temperature interval. Thus, the density and organization of domains in the adsorbed layer gradually change in a wide (more than 5 K) temperature interval, rather than exhibit a jump at $T_c$ (as it would be for the first-order phase transition).

\begin{figure}
\hspace{1in}
\epsfig{file=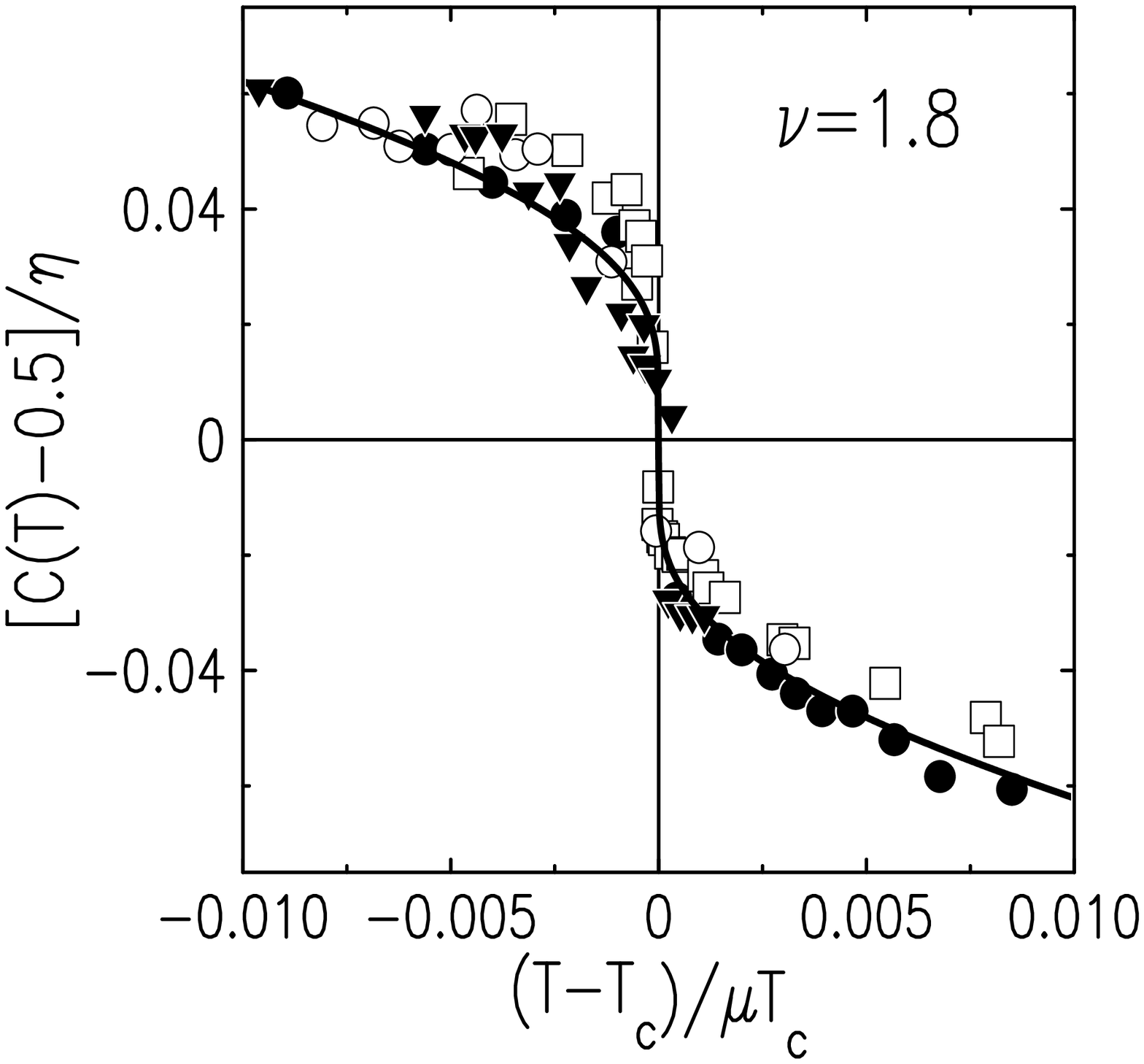, width=0.5\textwidth}

Figure 2. Temperature dependence of the surface fraction
$C(T)$ occupied by domains of the low-temperature condensed phase of alkanols
at the n-hexane–-water interface. Symbols represent experimental data [7, 8] for C$_{24}$OH
(filled circles) and C$_{30}$OH (open circles) and fluorinated alkanols FC$_{10}$OH (squares) and FC$_{12}$OH (triangles); the solid curve shows the results of calculations using Eq. (1)
with $\nu = 1.8$.
\end{figure}

The fractionation and intermixing of surface phases in the electric double layer at the liquid–-liquid
interface are possible due to competition between long- and short-range interactions [5, 10]. On the one hand, the electrostatic energy of the linear boundary between two phases with different surface polarizations exhibits a negative logarithmic singularity
(because its electric field decays with distance $r$
according to the $\sim 1/r$ law), which favors unlimited
intermixing of these phases. On the other hand, a short-range interaction of van der Waals nature determines the energy of linear tension $\gamma_l$ (i.e., the energy of
phase boundary formation with neglect of the logarithmic effect). The stabilization of the spatially inhomogeneous (fractal) structure on the surface in a certain vicinity of $T_c$, where the surface energy of the condensed phase is comparable with the energy of the gas
phase, takes place due to the finiteness of $\gamma_l>0$ (e.g.,
$\gamma_l\approx 10^{–11}$ N in a monolayer of FC$_{12}$OH [11]).

Marchenko [10] described the so-called $"$devil$'$s staircase$"$
of phase transitions with a point of accumulation
at $T_c$ in the system under consideration. In the limit of
a large number of phase transitions, surface fraction
$C(T)$ occupied by the domains of the low-temperature
condensed phase varies according to the following law:
\begin{equation}
C(T)-C(T_c)=\eta \cdot {\rm sign}\left(T_c-T\right)\left(\ln{\frac{\mu T_c}{|T_c-T|}}\right)
^{-\nu},
\end{equation}
where $T \to T_c$ and $\eta$, $\mu$ and $\nu$ are constant phenomenological quantities.
The identical equality $C(T_c) = 0.5$ determines transition temperature $T_c$. The product
$\mu T_c$ is a parameter that determines the region of existence of the spatially
inhomogeneous surface structure, $\eta$ characterizes the molecular properties of the system,
and exponent $\nu$ must be a universal parameter.

Figure 2 shows good agreement between the $C(T)$
dependencies determined in [7, 8], on the one hand,
and relationship (1) with the parameters listed in the
table, on the other hand. In contrast to estimations
obtained earlier [7, 8], this relationship between the
theoretical parameters of the critical crossover in
monolayers of alcohols at the n-hexane–-water interface allows all systems in which this phenomenon is
observed to be described using the same critical exponent $\nu = 1.8 \pm 0.4$. Then, $\mu$ is in fact the single parameter that can vary in relation (1), since the value of $\eta$ is
also the same (to within the given error) for all systems
under consideration. The latter circumstance is apparently due to the fact that the dipole moment of the molecules of all alcohols is determined primarily by
the presence of a hydroxy group.

In conclusion, it should be noted that relationship
(1) can also be used for description of the critical
crossover in Gibbs monolayers of alcohols during variation of the external pressure $p$ at
$T = \rm const$ [12, 13]. In this case, $C$ is considered as a function of external
pressure $p$.

The author is grateful to V.\,I.\,Marchenko for fruitful
discussions, in particular, for pointing out the necessity of using parameter $\mu$ in Eq. (1).

\small
\vspace{5mm}
Table 1. Parameters of critical crossover in monolayers of long-chain
alcohol molecules adsorbed at the n-hexane–-water interface
\vspace{2mm}

\begin{tabular}{|c|c|c|c|}
\hline
%&&&\\
Alconol &$T_c$(K)&$\eta$& $\mu$ \\
%&&&& \\
\hline
C$_{\rm 24}$OH &300.0  &  $8.0\pm  1.0$  &$1.0\pm 0.5$\\
%&&&& \\
C$_{\rm 30}$OH &302.5  &  $8.5\pm  0.9$  &$1.6\pm 0.6$\\
%&&&& \\
FC$_{\rm 10}$OH &300.7 &  $9.0\pm  1.0$  &$7.0\pm 1.0$\\
%&&&& \\
FC$_{\rm 12}$OH &314.0 &  $8.0\pm  1.0$  &$1.1\pm 0.5$\\
%&&&& \\
\hline
\end{tabular}

\vspace{2mm}

Critical exponent in Eq. (1) is $\nu = 1.8$. Data for FC$_{10}$OH
and C$_{30}$OH were obtained using the incoherent reflection
model, while those for FC$_{12}$OH and C$_{24}$OH were obtained
using the coherent reflection model. The error bars were estimated utilizing the conventional $\chi^2$-criteria at the confidence level 0.9.
\large

\end{document}